\documentclass[doc]{apa}
\usepackage{graphicx}
\usepackage{amsmath}

\begin{document}
\newcommand{\myvec}[1]{\ensuremath{\mathbf{#1} } }
\newcommand{\mat}[1]{\ensuremath{\mathbf{#1}}}
\newcommand{\Mlim}{\ensuremath{\mathbf{M}_{\rm{lim} } } }
\newcommand{\Mc}{\ensuremath{\mathbf{M^{o }} }}
\newcommand{\Mbar}{\ensuremath{\overline{\mathbf{M } } } }
\newcommand{\Hbar}{\ensuremath{\overline{\mathbf{H } } } }
\newcommand{\Sbar}{\ensuremath{\overline{\mathbf{S } } } }
\newcommand{\Mpre}{\ensuremath{\mathbf{M}_{\rm{pre} } } }

\newcommand{\IN}{\textrm{{\tiny IN}}}

\newcommand{\tin}[1]{\ensuremath{\myvec{t}^{\IN}_{#1 } } }
\newcommand{\cin}[1]{\ensuremath{\myvec{c}^{\IN}_{#1 } } }
\newcommand{\hin}[1]{\ensuremath{\myvec{h}^{\IN}_{#1 } } }
\newcommand{\ti}[1]{\ensuremath{\myvec{t}_{#1 } } }
\newcommand{\T}[1]{\ensuremath{\myvec{T}_{#1 } } }
\newcommand{\Psymb}[1]{\ensuremath{\myvec{P}_{#1 } } }
\newcommand{\tc}[1]{\ensuremath{\myvec{t}^{o}_{#1 } } } 
\newcommand{\h}[1]{\ensuremath{\myvec{h}_{#1 } } } 
\newcommand{\s}[1]{\ensuremath{\myvec{s}_{#1 } } } 
\newcommand{\F}[1]{\ensuremath{\myvec{f}_{#1 } } }
\newcommand{\ci}[1]{\ensuremath{\myvec{c}_{#1 }} }
\newcommand{\K}[1]{\ensuremath{\myvec{k}_{#1} } }

\newcommand{\tavg}[1]{\myvec{\underline{t}}_{#1} }
\newcommand{\proj}[1]{\ensuremath{\mat{P}_{#1} } }
\newcommand{\projcomp}[1]{\ensuremath{\tilde{\mat{P}}_{#1} } }
\newcommand{\Mtf}{\ensuremath{\mat{M}^{{\textrm{{\tiny TF}}}}}}
\newcommand{\Mft}{\ensuremath{\mat{C}}}
\newcommand{\MA}{\ensuremath{\mat{M}_{A} } }
\newcommand{\dM}{\ensuremath{\mat{\delta M} } }
\newcommand{\dS}{\ensuremath{\mat{\delta S} } }
\newcommand{\dMc}{\ensuremath{\mat{\delta \Mc} }}
\newcommand{\dMbar}{\ensuremath{\overline{\dM}}}
\newcommand{\dSbar}{\ensuremath{\overline{\dS}}}  
\newcommand{\dMcbar}{\ensuremath{\overline{\dM}^{o}}} 
\newcommand{\Tm}{\ensuremath{\mat{T}_{\rm{model}}}}

\newcommand{\rankiness}{\ensuremath{\mathfrak{R}}}

\newcommand{\bra}[1]{\ensuremath{\langle#1|}}

\newcommand{\ket}[1]{\ensuremath{|#1\rangle}}

\newcommand{\braket}[2]{\ensuremath{\langle#1|#2\rangle}}

\newcommand{\ketbra}[2]{\ensuremath{|#1\rangle\langle#2|}}

\newcommand{\aold}{\ensuremath{\alpha_{O}}}

\newcommand{\assoc}[2]{\ensuremath{{#1}\rightarrow{#2}}}
\newcommand{\scassoc}[2]{\textsc{#1}\ensuremath{\rightarrow}\textsc{#2}}
\newcommand{\INCM}{\ensuremath{I_{NCM}}}
\newcommand{\anew}{\ensuremath{\alpha_{N}}}
\renewcommand{\vec}[1]{\ensuremath{\mathbf{#1}}}
\newcommand{\vel}[1]{\ensuremath{\vec{v}^{#1}}}
\newcommand{\tvec}[1]{\ensuremath{\vec{t}^{#1}}}
\newcommand{\that}[1]{%
\ensuremath{%
		\hat{\vec{t}}^{#1}
}%
}\newcommand{\tdot}[2]{%
\ensuremath{%
		\vec{t}^{#1}\cdot\vec{t}^{#2}
}%
}
\newcommand{\tsub}[1]{%
\ensuremath{%
		\vec{t}_{#1}
}%
}
\newcommand{\tIN}[1]{%
\ensuremath{%
		\vec{t}^{\IN}_{#1}
}%
}
\newcommand{\cIN}[1]{%
\ensuremath{%
		\vec{c}^{IN}_{#1}
}%
}
\newcommand{\csub}[1]{%
\ensuremath{%
		\vec{c}_{#1}
}%
}
\newcommand{\hIN}[1]{%
\ensuremath{%
		\vec{h}^{IN}_{#1}
}%
}
\newcommand{\hsub}[1]{%
\ensuremath{%
		\vec{h}_{#1}
}%
}
\newcommand{\fsub}[1]{%
\ensuremath{%
		\vec{f}_{#1}
}%
}
\newcommand{\fIN}[1]{%
\ensuremath{%
		\vec{f}^{IN}_{#1}
}%
}

\newcommand{\set}[1]{%
	\ensuremath{%
	\mathfrak{#1}
	}%
}
\newcommand{\none}{
	\nonumber \\
}
\renewcommand{\vec}[1]{
\ensuremath{%
	\mathbf{#1}
}%
}

\newcommand{\MTF}{\ensuremath{\mat{M}^{\textrm{\tiny TF}}}}
\newcommand{\MFT}{\ensuremath{\mat{M}^{\textrm{\tiny FT}}}}

\newcommand{\myinsfig}[1]{%
	\refstepcounter{figure}	\label{myfig:#1} \addtocounter{figure}{-1}
	\ifapamode{	
		\includegraphics[width=0.9\textwidth]{figs/#1.eps}
	}{			
		\includegraphics[width=0.4\textwidth]{figs/#1.eps}
	}{			
		\includegraphics[width=0.6\textwidth]{figs/#1.eps}
	}
}
\newcommand{\myinssmfig}[1]{%
	\refstepcounter{figure}	\label{myfig:#1} \addtocounter{figure}{-1}
	\ifapamode{	
		\includegraphics[width=0.4\textwidth]{figs/#1.ps}
	}{			
		\includegraphics[width=0.15\textwidth]{figs/#1.ps}
	}{			
		\includegraphics[width=0.4\textwidth]{figs/#1.ps}
	}
}

\newcommand{\myinsrotfig}[1]{ 
	\refstepcounter{figure}	\label{myfig:#1} \addtocounter{figure}{-1}
	\ifapamode{ 
		\includegraphics[height=0.9\textheight]{figs/#1.eps}
	}{	
		~\hspace{0.05\textwidth}
		\includegraphics[angle=270,width=0.4\textwidth]{figs/#1.eps}
	}{ 	
		\includegraphics[angle=270,width=0.8\textwidth]{figs/#1.eps}
	}
}
\newcommand{\myinsbrotfig}[1]{ 
	\refstepcounter{figure}	\label{myfig:#1} \addtocounter{figure}{-1}
	\ifapamodeman{
		~\hspace{0.15\textwidth}
		\includegraphics[height=0.9\textheight]{figs/#1.ps}
	}{
		~\hspace{0.1\textwidth}
		\includegraphics[angle=270,width=0.8\textwidth]{figs/#1.ps}
	}
}

\newcommand{\Edit}[1]{
	{\large{\bf#1}}
}

\newcommand{\pair}[2]{\textsc{#1}-\textsc{#2}}

\newcommand{\Lk}{\ensuremath{\mathbf{L}^{\textrm{\scriptsize{-1}}}_{\textrm{k}}}}
\newcommand{\Tau}{\ensuremath{\overset{*}{\tau}}}
\newcommand{\taustar}{\ensuremath{\overset{*}{\tau}}}
\newcommand{\xstar}{\ensuremath{\overset{*}{x}}}
\newcommand{\pstar}{\ensuremath{\overset{*}{p}}}
\newcommand{\lstar}{\ensuremath{\overset{*}{p}}}
\newcommand{\Tauf}{\ensuremath{\overset{o}{\tau}}}
\newcommand{\po}{\mathcal{P}}
\newcommand{\To}{\mathcal{T}}
\newcommand{\Io}{\mathcal{I}}
\newcommand{\Ro}{\mathcal{R}}
\newcommand{\ftilde}{\ensuremath{\tilde{f}}}

\title{Evidence accumulation in a Laplace domain decision space}
\shorttitle{Evidence accumulation in the Laplace domain}
\leftheader{Evidence accumulation in the Laplace domain}
\rightheader{Evidence accumulation in the Laplace domain}

\author{Marc W.~Howard, Andre Luzardo, Zoran Tiganj}
\affiliation{Department of Psychological and Brain Sciences\\Department of
Physics\\Boston University}
\note{\today}

\abstract{ Evidence accumulation models of simple decision-making have long
		assumed that the brain estimates a scalar decision variable
		corresponding to the log-likelihood ratio of the two alternatives.
		Typical neural implementations of this algorithmic cognitive model
		assume that  large numbers of neurons are each noisy exemplars of the
		scalar decision variable. Here we propose a neural implementation of
		the diffusion model in which many neurons construct and maintain the
		Laplace transform of the distance to each of the decision bounds.
		As in classic findings from brain regions including LIP, the firing
		rate of neurons coding for the Laplace transform of net accumulated
		evidence grows to a bound during random dot motion tasks.  However,
		rather than noisy exemplars of a single mean value, this approach
		makes the novel prediction that firing rates grow to the bound
		exponentially; across neurons there should be a distribution of
		different rates.  A second set of neurons records an approximate
		inversion of the Laplace transform; these neurons directly estimate
		net accumulated evidence.  In analogy to time cells and place cells
		observed in the hippocampus and other brain regions, the neurons in
		this second set have receptive fields along a ``decision axis.'' This
		finding is consistent with recent findings from rodent recordings.
		This theoretical approach places simple evidence accumulation models
		in the same mathematical language as recent proposals for representing
		time and space in cognitive models for memory.  } 

	\acknowledgements{We acknowledge helpful discussions with Bing Brunton,
			Josh Gold, Chandramouli Chandrasekaran,  Chris Harvey, Ben Scott,
			and Karthik Shankar.  This work was supported by NIBIB
			R01EB022864, ONR MURI N00014-16-1-2832 and NIMH R01MH112169.
}	
		\maketitle{}

\newcommand{\Evidence}[1]{\ensuremath{\mathcal{E}_{#1}}}

\newlength{\figheight}
\setlength{\figheight}{0.2\textheight}

The computational models of cognition that are most influential  on
neuroscience were developed in mathematical psychology to account for behavior
without regard to neural constraints.  For instance, the \citeA{AtkiShif68}
model of working memory maintenance was developed to account for behavioral findings
from continuous paired associate learning and other behavioral memory tasks.
The central idea of the \citeA{AtkiShif68} model---that  working memory holds
a small number of recently-experienced stimuli in an activated state---went on
to be extremely influential on neurophysiological studies of working memory
maintenance \cite<e.g.,>{Gold96,FustJerv82} as well as computational
neuroscience models of working memory
\cite<e.g.,>{Gold09,LismIdia95,CompEtal00}.  Similarly, models for
reinforcement learning originally developed to account for behavior
\cite{SuttBart81} have been extremely influential in understanding the neural
basis of reward systems in the brain \cite{SchuEtal97,WaelEtal01}.
Mathematical models of evidence accumulation \cite{Lami68,Link75,Ratc78} have
organized and informed a large body of neurobiological work \cite{GoldShad07}.
There are many neurophysiological phenomena in each domain that do not follow
naturally from the neural predictions of these models---this is not surprising
insofar as they were developed in most cases long before any
relevant neurophysiological data was available.  Moreover, the models from
each domain seem very different from one another despite the fact that any
real-world behavior undoubtedly depends on interactions among essentially the
entire brain.  

Part of the mismatch perhaps follows from taking models developed at an
algorithmic level literally at an implementational or biological level.  In this
paper, we extend a formalism that has already been applied to behavioral and
neural data from working memory experiments \cite{HowaEtal15,TigaEtal18a} and
neural representations of space and time \cite{HowaEtal14} and show that it
applies also to models of evidence accumulation.  
This formalism estimates functions of variables out in the world by
constructing the Laplace transform of those functions and then inverting the
transform \cite{ShanHowa12,ShanHowa13,HowaEtal14}.  We show that this approach
leads to a neural implementation of the diffusion model \cite{Ratc78}.  This
neural implementation of the diffusion model  
provides an account of neural data that makes a number of novel quantitative
predictions.  

\subsection{The diffusion model and sequential sampling}

Models of response times have been a major focus of work in cognitive and
mathematical psychology for many decades \cite{Luce86}.  Evidence accumulation
models have received special attention \cite{Lami68,Link75,Ratc78}.   
These models hypothesize that during the time between the arrival of a probe
and the execution of a behavioral decision, the brain contains an internal variable the
dynamics of which describe progress towards the decision. 
For our purposes it will be
sufficient to describe the dynamics of accumulated evidence with the following
random walk:
\begin{equation}
		X_{t+\Delta} = X_{t} + \Delta \Evidence{t}
		\label{eq:DiffusionDef}
\end{equation}
where \Evidence{t} is the instantaneous evidence available at time $t$.
Equation~\ref{eq:DiffusionDef} describes a perfect integrator.
In simple evidence accumulation tasks,  
\Evidence{t} is not known in detail. Under these circumstances, one typically
takes \Evidence{t} to be a stochastic term. 

The diffusion model \cite{Ratc78} is  the most widely-used of the evidence
accumulation models.  In the diffusion model, one assumes \Evidence{t} is a 
normally distributed random variable with a non-zero mean referred to as the drift rate.  
The units of $X_t$  are usually chosen to be the standard deviation of the
normal distribution.  The response is emitted when $X_t$ reaches either of two
boundaries; the value at the lower and upper boundaries are referred to as $0$
and $a$ respectively.  In the diffusion model, the starting point of the
dynamics are referred to with a parameter $z$.

The diffusion model can be understood as an implementation of a sequential
ratio test, a normative solution to the problem
of forming a decision between two alternatives \cite{Wald45,Wald47,WaldWolf48}.
Suppose we have two
alternatives, \textsc{l} and \textsc{r} that could have generated a sequence
of independent observations of data, $d_i$.
Starting with a prior belief about the likelihood ratio of the two
alternatives $\frac{P_0(L)}{P_0(R)}$, we find that the
likelihood ratio of the two hypotheses after observing $t$ data points is 
\begin{equation}
		\frac{P(L | d_1 \ldots d_t)
			}{
				P(R| d_1 \ldots d_t)
		} = \prod_{i=1}^t \left[ \frac{P(d_i | L) } {P(d_i | R) }\right]  \frac{P_0(L)}{P_0(R)}
		\label{eq:seqdec}
\end{equation}
The likelihood ratio on the lhs tells us the degree of certainty
we have that the sequence of data was generated by alternative \textsc{l}; the
inverse of the likelihood ratio tells us the degree of certainty we have that
the data was generated by alternative \textsc{r}.   We can determine the level
of certainty we require to terminate sampling and execute a decision.

The logarithm of the likelihood ratio is closely related to the diffusion
model.
Taking the log of both sides of Eq.~\ref{eq:seqdec} we find  
\begin{equation}
		\log P(L | d_1 \ldots d_t)
		 - \log	
				P(R| d_1 \ldots d_t)
		=  \log P_0(L) - \log P_0(R) +  \sum_{i=1}^t \left[ \log P(d_i | L) -  \log P(d_i | R) \right]
		\label{eq:seqdeclog}
\end{equation}
Here the prior log  likelihood ratio appears as the starting point of a sum; each
additional observation contributes an additive change in the evidence for one
alternative over the other.  Reaching a criterion can be understood as the log
likelihood ratio reaching a particular threshold
(recall that $\log x = - \log(1/x)$).\footnote{\citeA{ZhanMalo12} provide an
outstanding discussion of the centrality of log likelihood to understanding
cognitive psychology.}  In this way the 
diffusion model can be understood as an implementation of a sequential
ratio test; the parameters of the diffusion model can be mapped onto
interpretable quantities.  The boundary separation $a$ can be understood as
proportional to the log likelihood ratio necessary to make a decision; the
starting point $z$ can be understood as a prior probability\footnote{When
$z=a/2$ the prior is uninformative.} and the drift rate can be understood as
the expectation of the log likelihood ratio contributed by each additional
sample.\footnote{In many experiments, such as the random dot motion task
discussed extensively below, it may be difficult to make a connection to the  
normative sequential sampling model. }

There are many variants on this basic strategy for evidence accumulation in
simple perception tasks that have been explored in the mathematical psychology
literature \cite<e.g.,>{UsheMcCl01,BogaEtal06,BrowHeat08}.  They have various
advantages and disadvantages but all retain the feature of dynamics of a
decision variable gradually growing towards a boundary of some type.   The
diffusion model and other models of this class have been used to measure behavioral
performance in an extremely wide range of behavioral tasks in humans
\cite<e.g.,>{RatcMcKo08}.

\subsection{Neural evidence for evidence accumulation models}

\begin{figure*}
		\centering
		\begin{tabular}{lclc}
				\textbf{a} && \textbf{b}\\
				&\includegraphics[height=0.3\textheight]{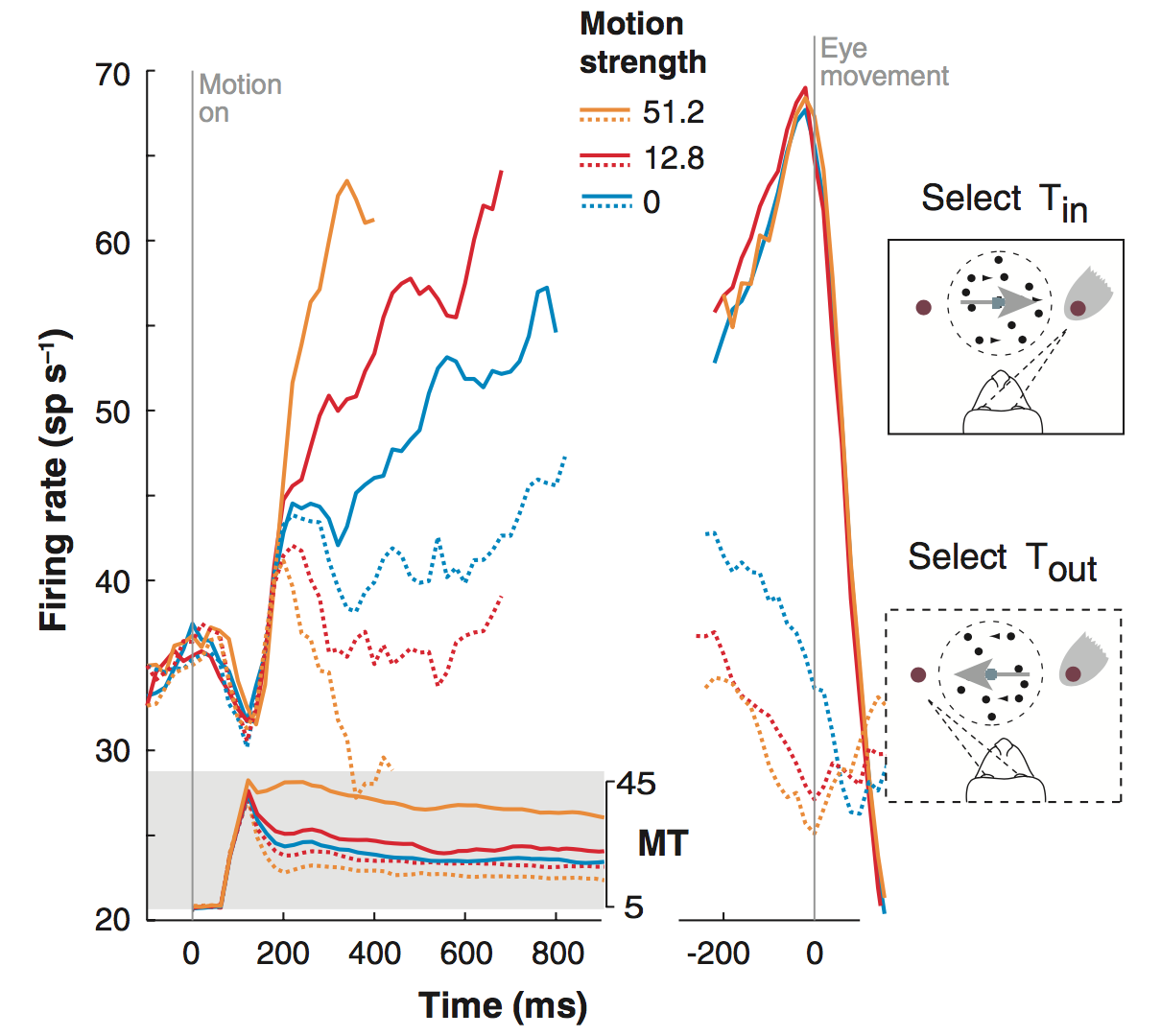}
				&&\includegraphics[height=0.3\textheight]{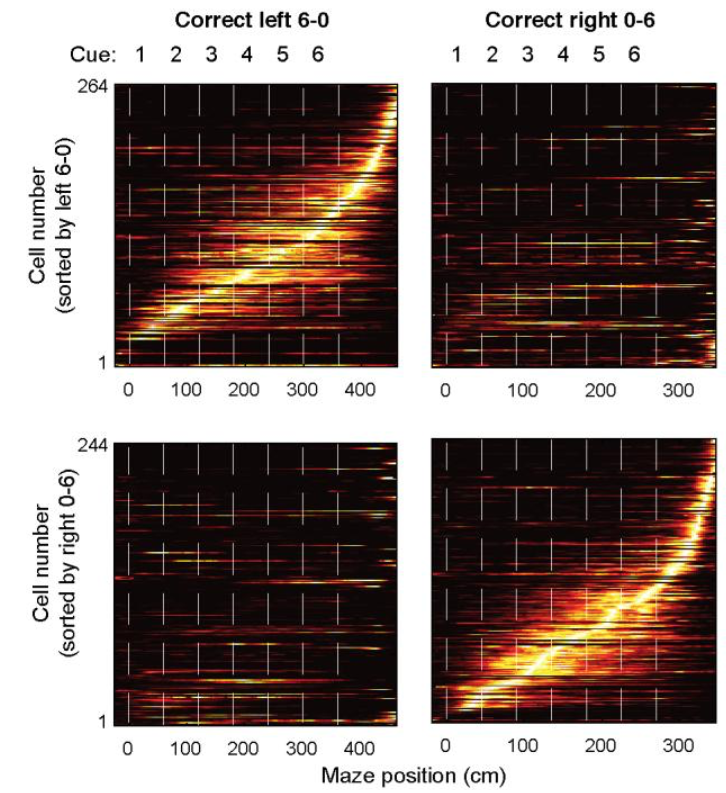}
		\end{tabular}
		\caption{\textbf{Neural evidence for evidence accumulation.}
		\textbf{a.} In some brain regions, neurons accumulate evidence towards a
			bound.  These studies typically use random dot motion.  The plot
			shows average firing rate across a number of LIP neurons.
			When the decision was made in the cells' preferred direction (solid
			lines) the firing rate grew as evidence accumulated, with more
			rapid accumulation for greater degrees of coherence.  When the
			response was in the other direction, firing rate decreased.  When
			aligned on the response (right), the neurons' firing rate was
			tightly coupled at the time of response, as if the response was
			triggered when the firing of the population reached a threshold.
			After Gold \& Shadlen (2007). \textbf{b.} In
			other brain regions (e.g., PPC), the firing rate depends on
			position along a decision path.   In this experiment, the animal
			ran down a corridor in virtual reality.  Visual stimuli were
			presented on the left or the right; at the end of the corridor,
			the animal was rewarded for turning in the direction that had more
			stimuli.  Each row represents the
			firing rate of a neuron as a function of position along the track.
			The left panels show trials on which all of the stimuli were on
			the left; the right panels show trials on which all of the stimuli
			were on the right.  The top panels show cells that were selective
			for left turn trials; the bottom panels show cells that were
			selective for right turn trials.  After Morcos \& Harvey (2016).
		}
		\label{fig:Neural}
\end{figure*}

There are two broad classes of evidence from the neurobiology of evidence
accumulation that we will review here.  First, there is evidence dating from
the late 1990's for neurons
whose firing rate appears to integrate information to a bound.  These studies
are typically done in monkeys in the random-dot-motion paradigm.  Second, a
more recent body of work shows evidence for neurons that activate
heterogeneously and sequentially during information integration.  These
studies typically use rodents in experimental paradigms where the evidence
to be accumulated is under more precise temporal control. These literatures differ not
only in species, recording techniques and behavioral task, but also in the
brain regions that are investigated.  The theoretical framework we will present
provides a possible link between these domains, with neurons that integrate to
a bound corresponding to the Laplace transform of the function describing
distance to the decision bound and neurons that activate
sequentially corresponding to the inverse transform of this function.

\subsubsection{Neurons that integrate to a bound}

Models of evidence accumulation have been used to understand the
neurobiology of simple perceptual judgments.  In a pioneering study,
\citeA{HaneScha96} found that the firing rate of neurons in the frontal eye
fields (FEF) predicted the time of a movement in a voluntary movement
initiation experiment.  During the preparatory period, the firing rate grew.
When the firing rate reached a particular value, the movement was initiated
and could not be terminated via an instruction to terminate the movement.  In
subsequent studies, neurons in the lateral interparietal area (LIP) appeared
to comport with the characteristic predictions of evidence accumulation models
in a simple decision-making task.  

In the random-dot-motion (RDM) paradigm, a display consisting of moving dots
appears in the visual field.  Some proportion of dots move in the same
direction; a movement (typically a saccade) in the direction of the coherent
movement is rewarded.  The decision made by monkeys is predicted above chance
by the  particular sequence of random dots even though there is on average no
coherent movement \cite{KianEtal08}.  Neurons in the middle temporal area (MT)
fire in response to these moving stimuli and predict perceptual
discriminability in the task \cite{NewsEtal89,BritEtal92}.  During performance
of the random dot motion task, the firing rate of LIP neurons with fields in
the  movement direction grows during the decision-making period
\cite{ShadNews01}.  Conversely, neurons with receptive fields in the other
direction show a decrease in firing rate.  
Similar results have been found in a number of brain regions
\cite<see>[for a review]{BrodHank16}
and LIP is not required for decisions \cite{KatzEtal16}.
For neurons with receptive fields
corresponding to the correct direction, firing rate grows faster when the
motion coherence is greater and their firing rate is approximately constant
around the time at which a response is made (\citeNP{RoitShad02}, see also
\cite{CookMaun02a} for similar results in ventral intraparietal cortex, VIP).
Figure~\ref{fig:Neural}a summarizes many of these findings \cite{GoldShad07}.

A number of computational neuroscientists and cognitive modelers have
understood the neurons as indexing a decision-variable that changes as
additional evidence is accumulated \cite<e.g.,>{SmitRatc04,Wang08,BeckEtal08}.
The standard assumption in these approaches has been that a large population
of neurons each provides a noisy estimate of the instantaneous evidence.  Each
neuron signals the decision variable \emph{via} its firing rate; by averaging
over many neurons one can compute a better estimate of the magnitude of the
decision variable \cite<e.g.,>{ZandEtal14}. 

\subsubsection{Sequential neural responses during evidence accumulation}

However, recent evidence suggests that rather than many neurons being noisy
exemplars of a single scalar strength, there is in many brain regions
heterogeneity in the response of units in simple decision-making tasks
\cite{ScotEtal17,MorcHarv16,HankEtal15,MeisEtal13}.  In these studies, rather than the
RDM paradigm, the task allows more precise control over the time at which
evidence becomes available, enabling a detailed investigation of the effect of
information at different times on behavior  \cite{BrunEtal13} and also neural
responses.  For instance, one might present a series of clicks on one side of
an animal's head or the other and reward a response to the side that had more
clicks \cite{BrunEtal13,HankEtal15}.  Other variants of this approach can
utilize flashes of light \cite{ScotEtal17} or visual stimuli presented along
the left or right side of a corridor in virtual reality \cite{MorcHarv16}.  

The critical result from these studies is that rather than changing firing
rate monotonically with the decision variable as evidence is accumulated, in
several experiments neurons instead respond in a sequence as the decision
variable changes.  The experiment of \citeA{MorcHarv16} provides a very clear
result.  In this experiment, mice ran along a
virtual corridor.  During the run, a series of distinctive visual stimuli were
presented along the left or the right side of the virtual corridor.  At the
end of the corridor, the mice were rewarded for turning in the direction that
had more visual stimuli.   \citeA{MorcHarv16} observed that neurons in the
posterior parietal cortex (PPC) were activated as if they had receptive fields
along a decision axis.  Figure~\ref{fig:Neural}b  illustrates the key findings
of \citeA{MorcHarv16}.  On trials where all of the evidence was in one
direction or the other, the population fired sequentially, with distinct
sequences for progress towards the different decisions.  Notably, the neurons
in Figure~\ref{fig:Neural}b seem to show an overrepresentation of points near
the decision bound.  This can be seen from the ``hook'' in
Figure~\ref{fig:Neural}b.

\subsection{The Laplace transform in cognitive science and computational
neuroscience}

In this paper we pursue the hypothesis that the two classes of neural data from
evidence accumulation experiments can be understood as the Laplace transform
and inverse transform of a function describing the net evidence since the
decision-making period began.  This places models of evidence accumulation in
the same theoretical framework as recent models of memory that utilize the
Laplace transform and its inverse \cite{ShanHowa12,ShanHowa13} to construct
cognitive models of a range of memory tasks \cite{HowaEtal15}.  The Laplace transform
can also be used to construct neural models of time cells and place cells
\cite{HowaEtal14}. 

The cooperative behavior of many neurons can be understood as representing
functions over  variables in the world.  For instance, individual photoreceptors respond
to light in a small region of the retina.  The activity of a set of many
photoreceptors can be understood as representing  the pattern of light as a
function of location on the retina.  For some variables, such as the location
of light on the retina or the frequency of a sound, the problem of how
to represent functions amounts to the problem of placing receptors in the
correct location along a spatial gradient (e.g., hair cells at different
positions along the cochlea are stimulated by different frequencies).  For
other variables, such as time and allocentric position, we cannot simply place
receptors to directly detect the function of interest.  For instance, ``time
cells'' observed in a range of brain regions appear to code for the time since
a relevant stimulus was experienced
\cite{MacDEtal11,MellEtal15,BolkEtal17,TigaEtal18a}.  The stimulus that is
being represented is in the past (in some experiments as much as one minute in
the past).  Similarly, (at least in some circumstances) place cells in the
hippocampus represent distance from an environmental landmark
\cite{GothEtal96}.  These cells show the same form of activity in the dark
\cite{GothEtal01} and cannot reflect locally available cues.  How can the
brain construct functions of time and space? One solution is that the brain
does not directly estimate the function of interest.  Rather the brain
maintains the Laplace transform of the function of interest and then inverts
the transform to estimate the function directly.   

\setlength{\figheight}{0.18\textheight}
\begin{figure*}
		\begin{tabular}{lclclc}
				\textbf{a} &&\textbf{b}&&\textbf{c}\\
				&
				\includegraphics[height=\figheight]{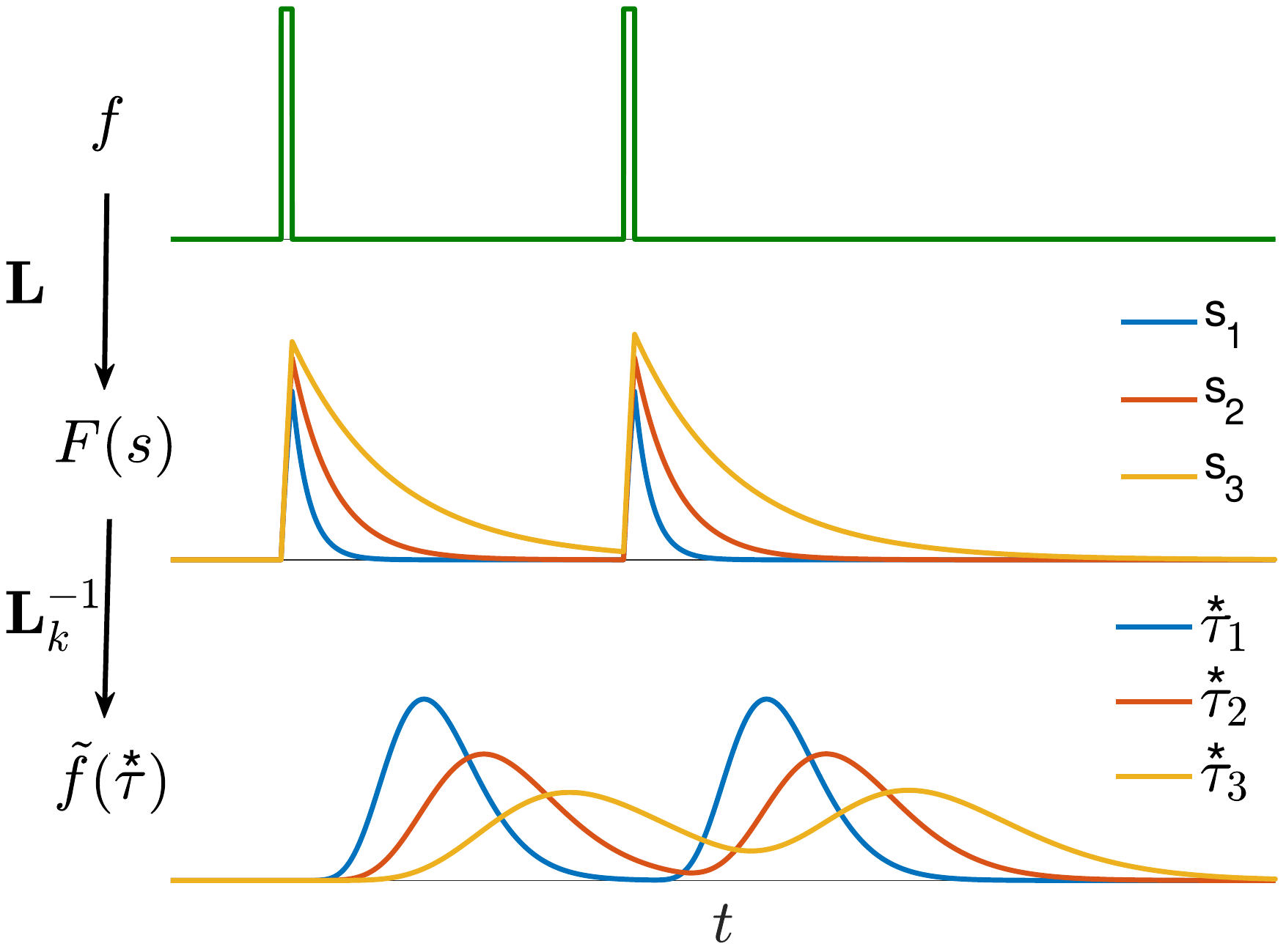}
				&&
				\includegraphics[height=\figheight]{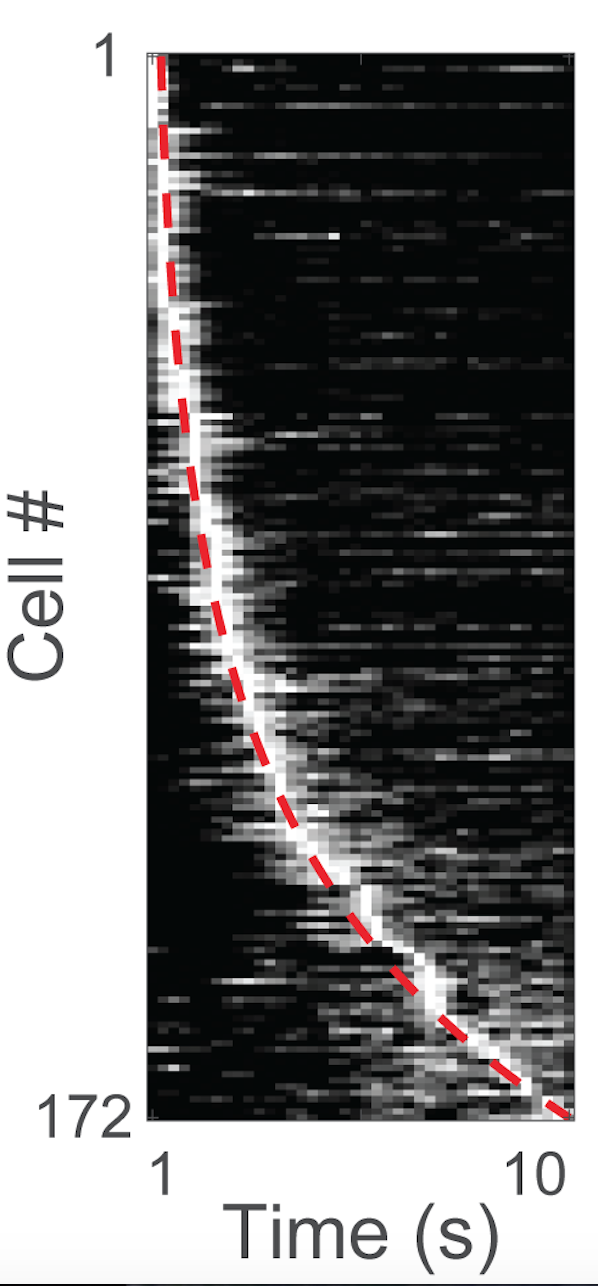}
				&&
				\includegraphics[height=1.1\figheight]{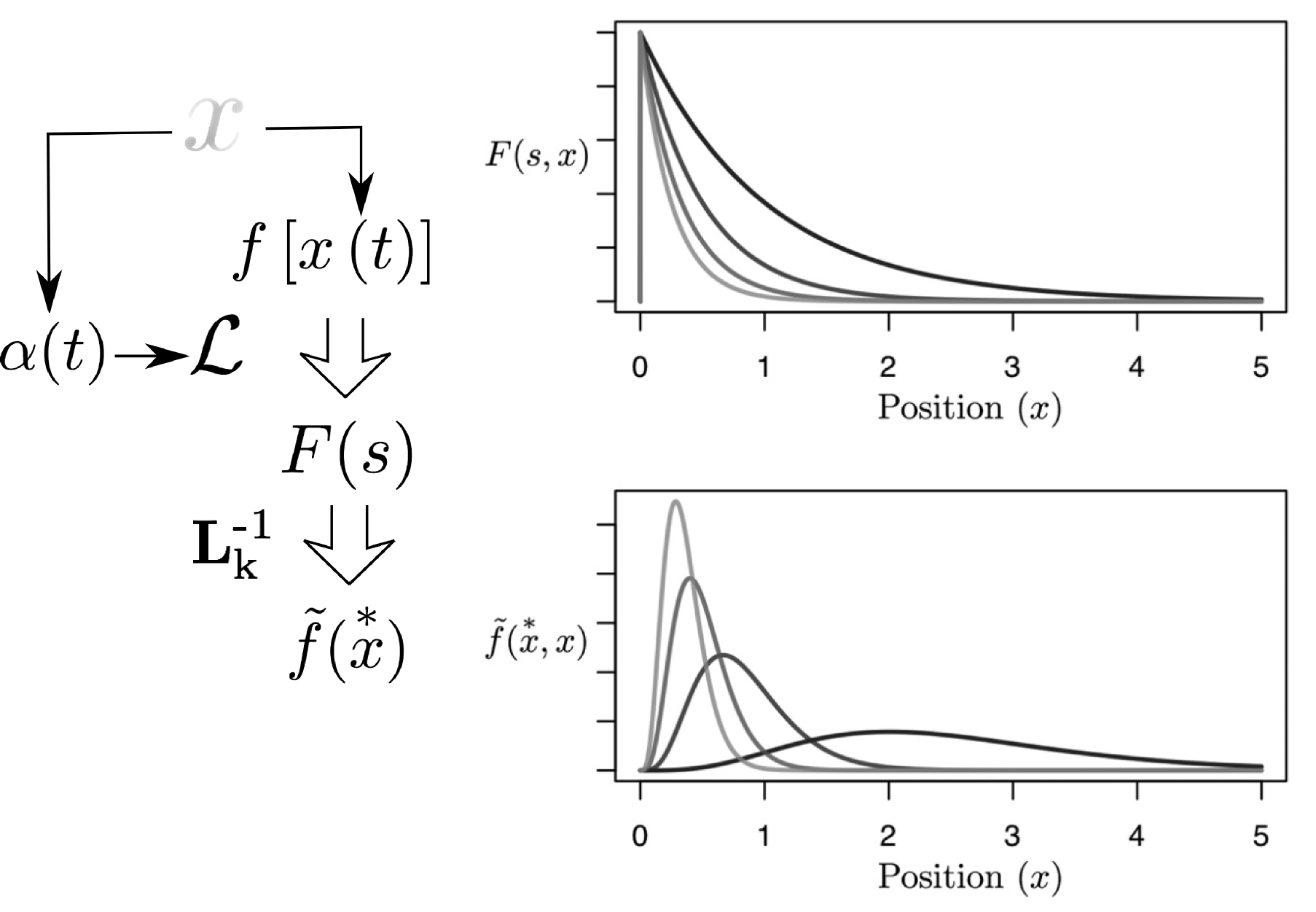}
		\end{tabular}
		\caption{
				The Laplace transform for time and space.
				\textbf{a.} The Laplace transform of functions of time.
				Consider an input $f(t)$ that is non-zero for a brief period.
				After the input, the different neurons corresponding to the
				Laplace transform, $F(s)$ activate and then decay
				exponentially.  Different neurons have different values of $s$
				and thus decay at different rates.  Neurons with these
				properties have recently been observed in the lateral
				entorhinal cortex (Tsao, et al., 2018).  Neurons representing
				the inverse transform $\ftilde(\taustar)$ respond a
				characteristic time after the input.  They fire when the past
				stimulus enters their receptive fields, defined by their value
				of $\taustar$. 
				\textbf{b.} Neural functions of time.  If the
				brain contained sets of neurons that represented functions of
				time, we would see sequentially-activated neurons.  Here we
				see simultaneously-recorded sequentially activated neurons in
				the hippocampus recorded using calcium imaging.  Each row
				gives the activation of  one neuron as a function of time
				during the delay.  The neurons are sorted according to their
				peak time.   Note the curvature.  This means that there are
				progressively fewer neurons coding for later in the delay.
				The dashed red line gives an analytic curve under the
				assumption of logarithmic compression.  Although this
				recording technique does not allow one to measure the width of
				individual cells' tuning, many papers using extracellular
				electrophysiology have shown that time fields also grow wider
				as the sequence unfolds. After Mau, et al., (2018).
				\textbf{c.} Laplace transform for variables other than time.
				Left: Schematic for the method for constructing the Laplace
				transform of variables $x$ in the world.  With access to the
				time derivative of $x$, the method modifies the differential
				equation for maintaining the Laplace tranform of time by
				$\alpha(t) = \frac{dx}{dt}$.  If the input $f$ is a function
				of $x(t)$, then $F(s)$ maintains the Laplace transform of
				$f(x)$.  Right: Consider a case in which the animal encounters
				a spatial landmark at $x=0$ causing an input \emph{via} $f$ and then
				moves in the neighborhood of the landmark (with $x>0$).  As
				the animal moves away, $dx/dt > 0$ and the cells in $F(s)$
				decay exponentially.
				But if the animal turns around and heads back toward the
				startng point, $dx/dt < 0$ and the cells grow exponentially.
				However, firing is always just a function of $x$.   Under
				these circumstances, the Laplace transform behaves like border
				cells with different space constants and neurons participating
				in the inverse Laplace transform behave like place cells.
				After Howard, et al., (2014).
				\label{fig:LaplaceNeural}
		}
\end{figure*}

\nocite{MauEtal18}
\nocite{TigaEtal18a}

\subsubsection{The Laplace transform  contains all
the information about the transformed function}
The mathematics of the Laplace transform are well-understood.  Given a
function $f(t)$, the Laplace transform of $f(t)$
is written as $F(s)$:
\begin{equation}
		F(s) = \int_0^\infty e^{-st'} f(t') dt'
		\label{eq:LaplaceDef}
\end{equation}
That is, one starts with a function  $f(t)$ that specifies a numerical value
at many different values of $t$.   By computing $F(s)$ via Eq.~\ref{eq:LaplaceDef} with
many different values of $s$,
we construct the Laplace transform of $f$.\footnote{For our purposes it is
sufficient to consider real values of $s$. }  Each value of $s$ simply gives
the sum of the product of $f(t)$ with an exponentially-decaying kernel that peaks at
zero. The Laplace transform has been studied for hundreds of years and is
widely-used in engineering applications.  

The most important property of the Laplace transform for present purposes is
that the transfrom can be
inverted.  Put simply, that means that if we know the values of $F(s)$ for
each value of $s$, we can in principle recover the value of $f(t)$ at every
value of $t$.  The most widely-used algorithms for inverting the transform
involve taking the limit of an integral computed in the complex plane.  These
methods are complicated and difficult to implement neurally.
Fortuitously, there is an algorithm for inverting the Laplace transform that
only requires real coefficients and nothing more demanding than computing
derivatives \cite{Post30}.  This technique, referred to as the Post
approximation, yields a scale-invariant approximation of the inverse
transform, and thus recovers the original function \cite{ShanHowa12} and is neurally
realistic \cite{ShanHowa12,LiuEtal18}.

Formally, we can keep track of the Laplace transform with a differential equation:
\begin{equation}
		\frac{dF(s)}{dt} =  -s F\left(s\right) + f\left(t\right) 
		\label{eq:Laplacenoalpha}
\end{equation}
The solution to this equation at time $t$:
\begin{equation}
		F_t\left(s\right) = \int_{-\infty}^t e^{-s\left(t-t'\right)} f(t') dt'
		\label{eq:Laplacesolve}
\end{equation}
is just the Laplace transform of $f(t' < t)$.  Note that updating
Eq.~\ref{eq:Laplacenoalpha} as time unfolds requires only information about
the input available at time $t$ and the value of $F(s)$ at the immediately
preceding moment.  
Because Eq.~\ref{eq:Laplacenoalpha} implements the Laplace transform of
$f(t'<t)$, we know that a set of units $F_t(s)$ contains all of the
information present in the function $f(t' < t)$.  Because the function is the input
presented over the past, we can conclude that at time $t$, $F(s)$ maintains
the Laplace transform of the past.  If we could invert the transform, we could
recover the function of the past itself.

The Post approximation provides a way to approximately invert the transform.
This method can be written as follows:
\begin{eqnarray}
		\ftilde\left(\taustar\right) &\equiv & \Lk F(s) 
		\label{eq:ftilde}
		\\
						&=& C_k s^{k+1} \frac{d^k}{ds^k} F(s)
\end{eqnarray}
where $\taustar \equiv -k/s$,  $k$ is an integer constant that controls the precision of the inverse,
$C_k$ is a constant that depends on $k$ and $\frac{d^k}{ds^k}$ means to take 
the $k$th derivative with respect to $s$.   \citeA{Post30} proved that in the
limit as $k\rightarrow \infty$, the inverse is precise and
$\ftilde_t(\taustar) = f(t+\taustar)$.  The variable $\taustar$ is negative;
for each unit in $\ftilde$
its value of $\taustar$ characterizes the time in the past that unit
represents.  When $k$ is small, this method yields only an approximate
estimate of the function.  The errors in the reconstruction appear as
receptive fields in time with a width that increases for time points further
in the past.   Figure~\ref{fig:LaplaceNeural}a illustrates the time dynamics
of $F(s)$ and $\ftilde(\taustar)$ for a simple input.  It is worth noting that
the time dynamics of $F(s)$ and $\ftilde(\taustar)$ resemble
neurophysiological findings of so-called ``temporal context cells'' in the
lateral entorhinal cortex  \cite{TsaoEtal18} and ``time cells'' observed in
the hippocampus and other regions \cite{PastEtal08,MacDEtal11,MauEtal18} respectively.

\subsubsection{Generalization to hidden variables other than time}
Equations~\ref{eq:Laplacenoalpha}~and~\ref{eq:ftilde} provide a concise account
for representing functions of time \cite{ShanHowa12} that mimics the firing of
time cells in a range of brain regions \cite<Fig.~\ref{fig:LaplaceNeural}b,
see also>{MauEtal18,TigaEtal18a} and can be
readily implemented in realistic neural models \cite{TigaEtal15,LiuEtal18}.
In this paper we exploit the fact that this coding scheme can also be used to
construct functions over variables other than time.  Consider the simple
generalization of Eq.~\ref{eq:Laplacenoalpha}:
\begin{equation}
		\frac{dF(s)}{dt} = \alpha(t)\left[ -s F\left(s\right) + f\left(t\right) \right]
		\label{eq:Laplacealpha}
\end{equation}
Note that this reduces to Eq.~\ref{eq:Laplacenoalpha} if $\alpha(t) = 1$.
If one can arrange for $\alpha(t)$ to be equal to the time derivative of some
variable $x$, $\alpha(t) = \frac{dx}{dt}$, then one can use
Eq.~\ref{eq:Laplacealpha} to construct the Laplace transform of functions over $x$ instead of over time
\cite{HowaEtal14}.\footnote{to see this, set $\alpha(t) = \frac{dx}{dt}$ and multiply
both sides of Eq.~\ref{eq:Laplacealpha} by $\frac{dt}{dx}$.}  For instance, if
$f(t)$ is non-zero only when an animal encounters the start box of a linear
track, and if $\alpha(t)$ is the animal's
velocity along the track, then as the animal runs back and forth along the
track, $\alpha(t)$ changes sign. During these periods of time, the firing rate
of each unit in
$F(s)$ decays as an exponentially-decreasing function of distance.  At the same
time, units in $\ftilde$ behave like one-dimensional place cells
(Figure~\ref{fig:LaplaceNeural}c).  When the 
inverse is understandable as a function over some variable other than time, we 
will write $\ftilde(\xstar)$.

\subsubsection{Optimal Weber-Fechner distribution of $s$}

Thus far we've discussed at a formal level how to use many neurons with
different values of $s$ to represent the Laplace transform of functions and
how to use the operator $\Lk$ to construct an approximation
of the original function $\ftilde(\xstar)$ with many neurons corresponding to
many values of $\xstar$.  It remains to determine how to allocate neurons to
values of $s$ and $\xstar$.  Because $\xstar$ is in one-to-one relationship
with $s$, it is sufficient to specify the allocation of neurons to  $s$.  

It has been argued that it is optimal to allocate neurons to represent a
continuous variable in such a way that receptive fields are evenly spaced as a
function of the logarithm of that variable \cite{HowaShan18}.  In addition
to enabling a natural explanation of the Weber-Fechner law, positioning
receptors evenly along a  logarithmic scale also enables the neural system to
extract the same amount of information from functions with a wide range of
intrinsic scales.  This logarithmic scaling requires that receptive field
center of the $n$th receptor goes up like $\xstar_n = c^n$.\footnote{This
		implies the ordinal variable $n \propto \log
\xstar$.} This compression results in a characteristic ``hook'' in the
heatmaps constructed by sorting neurons on their peak time (as in
Fig.~\ref{fig:LaplaceNeural}b).  Coupled with a linear increase in receptive field
width with an increase in $x$,\footnote{An increase in receptive field width
would appear as increase in the width of the central ridge in
Fig.~\ref{fig:LaplaceNeural}b.  This spread is not visible due to the
properties of the recording method used in that study; an increase in
receptive field width with peak time is observed in time cell studies using
other recording techniques
\cite{JinEtal09,SalzEtal16,MellEtal15,TigaEtal18a}.}
this arrangement means that the acuity between adjacent receptors is a
constant.

\section{The  Laplace transform of the Diffusion Model}
\begin{figure*}
		\centering
		\includegraphics[width=0.8\textwidth]{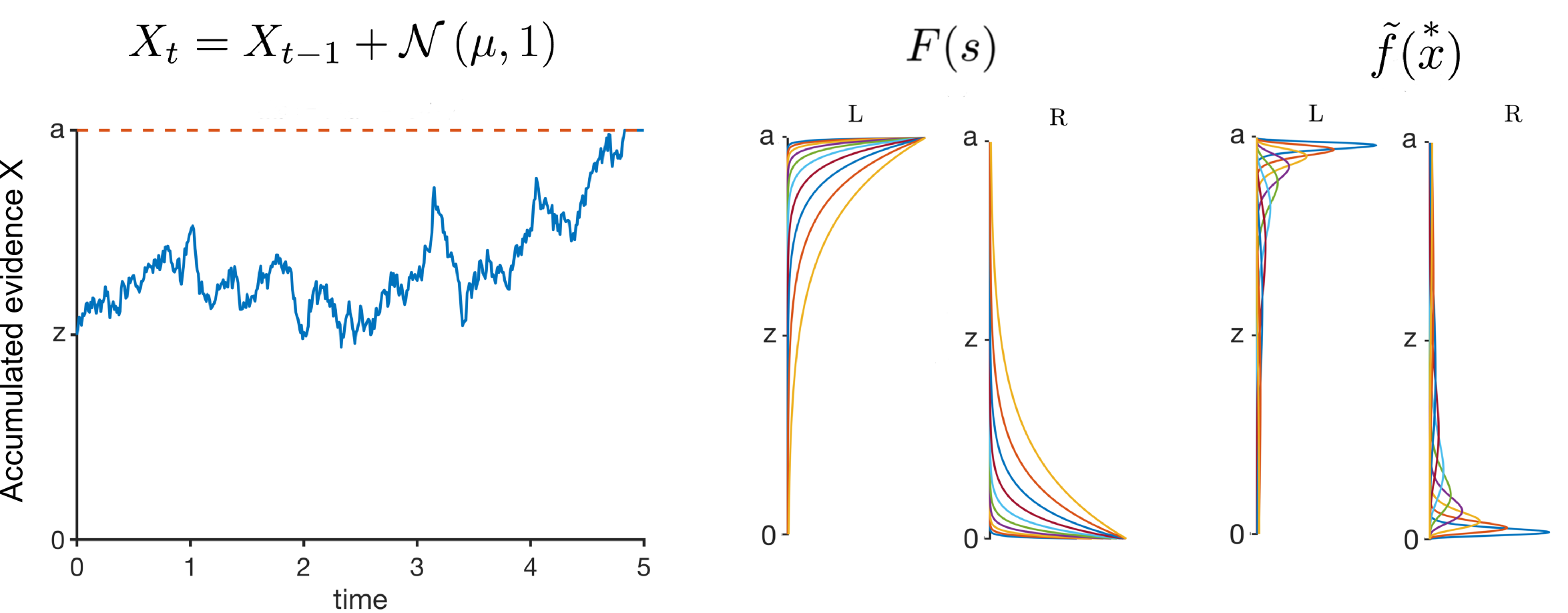}
		\caption{Implementing the diffusion model with many neurons \emph{via}
		the Laplace transform.  At an algorithmic level, the diffusion model
		describes an abstract  particle
		representing the quantity of net accumulated evidence moving towards
		two boundaries.  At the
		implementational level, the position of this particle is represented
		in the Laplace domain by two sets of units that code for the Laplace
		transform of position relative to the two boundaries, $F(s)$.  These
		units have exponentially-decaying ``receptive fields'' with respect to
		net evidence.  Each 
		unit is parameterized by a value $s$ that controls its rate of
		decay---different colors correspond to different values of $s$. 
		Another set of units $\ftilde(\xstar)$ approximates the inverse
		Laplace transform.  Units in this representation have receptive fields
		that tile the evidence axis.  Each unit is parameterized by a value
		$\xstar$ that controls the center of its receptive field.  These units
		are compressed and scale-invariant in the same way that time cells
		generated by the same equations are (Figure~\ref{fig:LaplaceNeural}).
		}
		\label{fig:DMschem}
\end{figure*}

We propose a model of evidence accumulation in which the Laplace transform of
accumulated evidence is represented at each moment and inverted
\emph{via} a linear operator.  The result is evidence on a supported dimension
with logarithmic compression.
The diffusion model describes the position of a particle relative to two
decision boundaries.  Inspired by the curvature present in the
\citeA{MorcHarv16} data, we  maintain the Laplace transform 
of the distance to each of the two boundaries, treating the appropriate
decision bound as zero for each of the two functions (Figure~\ref{fig:DMschem}).  This formalism provides
a mapping between the diffusion model at an algorithmic level and neurons at
an implementational level.

Let us refer to the two alternative responses as \textsc{l}  and \textsc{r}
and the net evidence for the left alternative accumulated up to time $t$ as
$L(t) - R(t)$.  The decision-maker's goal is to estimate the net
evidence that has been experienced since the trial began.  The instantaneous
evidence gives the time derivative of that variable.   
In the random dot motion paradigm we would expect the instantaneous derivative
to fluctuate  more or less  continuously; the mean of this derivative would
differ across conditions with different levels of motion coherence.  In
paradigms with discrete clicks or flashes of light, the presence of a click
signals a non-zero value of the derivative, with the side of the click
controlling the sign of the derivative.

We start constructing this model by assuming that we have two sets of leaky
accumulators $F_L(s)$ and $F_R(s)$.  
These correspond to opponent pairs of
evidence accumulators.
$F_L(s)$  codes for the Laplace transform of $L-R$ over the range $0$ to $a$
and $F_R(s)$ codes for the Laplace transform of $R-L$ over the domain $a$ to
$0$.  
We adopt the strategy of assigning each $\alpha(t)$ to the derivative of this
decision variable; $\alpha_L(t) = - \alpha_R(t)$.
When evidence accumulation begins, we want to have  
\begin{eqnarray}
		F_L(s,0) & = & e^{-sz} \nonumber \\
		F_R(s,0) & = & e^{-s(a-z)} 
		\label{eq:Init}
\end{eqnarray}
where $z$ corresponds to the bias term and $a$ corresponds to the boundary
separation.
Note that these correspond to the same location on the decision axis, but
reference to different starting points.  This initialization can be
accomplished by ``pulsing'' each accumulator with a large $\alpha$ value. The
sum of the area under these two pulses  controls the boundary separation.  The
difference controls the bias.

After initialization, as evidence is accumulated, we set $\alpha_L(t) = \frac{d(L-R)}{dt}$ 
and $\alpha_R(t) = \frac{d(R-L)}{dt}$ and evolve both $F_L(s)$ and $F_R(s)$
according to Eq.~\ref{eq:Laplacealpha}.  In this way each set of integrators codes for
the Laplace transform of the distance to each of the two decision boundaries.  

At each moment a second set of units holds the approximate inversion of the
Laplace transform, $\ftilde_L(\xstar)$ and $\ftilde_R(\xstar)$ using
\begin{eqnarray}
		\ftilde_L(\xstar) &=& \Lk F_L(s) \nonumber \\
		\ftilde_R(\xstar) &=& \Lk F_R(s) 
		\label{eq:LRftilde}
\end{eqnarray}
By analogy to time cells and place cells, these units tile the decision space
with compressed receptive fields that grow sharper in precision as each unit's
preferred decision outcome is approached.

\subsection{Compression of $\xstar$ gives equal discriminability
in probability space}
Noting the characteristic curvature in the \citeA{MorcHarv16} data
(Fig.~\ref{fig:Neural}), we configure each set of integrators to represent
distance to the appropriate decision bound.  
Following previous work on time and space \cite{HowaShan18}, we choose a
Weber-Fechner scaling of the $\xstar$ axis.  Denoting the value of $\xstar$
for each unit like, $\xstar_1, \xstar_2
\ldots \xstar_n$ gives
$\xstar_i \propto C^n$, where $C$ is some constant that depends
on the number of units and
their relative precision \cite{HowaShan18}.  
Because of the properties of the inverse Laplace transform, the width of
each receptive field goes up proportional to the unit's value of $\xstar$.
From this expression, we can see that units are evenly spaced on a logarithmic
scale.  

These properties imply that
the set of units has greater discriminability for smaller values of $\xstar$;
as $\xstar$ increases the spacing between adjacent units increases. 
In the context of sequential ratio testing, the distance to the bound in the
diffusion model can be understood as the distance to the bound for the log
likelihood ratio. 
The logarithmic spacing of $\xstar$ means that discriminability is equivalent
in units of \emph{likelihood ratio}.   This is a unique prediction of this
approach that may have important theoretical and empirical implications.

\subsection{Implementing response bias and imposing a response deadline}

This neural circuit is externally controllable \emph{via} $\alpha(t)$.  
Note that $\alpha(t)$ affects each unit in a way that is appropriate to its
value of $s$. 
It can be shown that affecting $\alpha(t)$ for a single cell is equivalent to
changing the slope of the f-i curve relating firing rate to the magnitude of an
internal current \cite{LiuEtal18}.  Changing $\alpha(t)$ for all the units in
$F(s)$ simply means to change the slope of all of their f-i curves
appropriately.  There are many biologically plausible mechanisms to rapidly change the
slope of the f-i curve of individual neurons
\cite<e.g.,>{ChanEtal02,Silv10}.
Because we can understand the implications of manipulating $\alpha(t)$ in
terms of the entire function $F(s)$, this lets
us readily determine conditions to implement response bias and change the
boundary separation to comply with a response deadline or even
a continuously-changing estimate of the cost of further deliberation
\cite{GersEtal15}.  

In order to initialize a response bias as in Eq.~\ref{eq:Init}, starting from
zero activation for all the units in both sets of $F(s)$, a pulse in $\alpha$
will push each set of units away from their respective bound.  The sum of the
pulses across the two accumulators is interpretable as the boundary
separation.    Response bias can be simply implemented by providing a
different magnitude pulse to the two sets of accumulators.  If each set of
units is given the same magnitude of a pulse in $\alpha$, there is no response
bias.  The accumulator that receives the smaller pulse starts evidence
accumulation closer to the bound. 

Similar mechanisms can be used to change the effective distance to the
response boundary rapidly. In the simulations below, we assume that the
appropriate response is executed when the activation of the first entry in
$\ftilde$,  $\ftilde(\xstar_1)$ reaches a threshold.  If $\ftilde_L(\xstar_1)$
is activated before $\ftilde_R(\xstar_1)$, option \textsc{l} is selected.  In
the same way that $\alpha_{L/R}(t)$ can be used to initialize evidence
accumulation by pushing  $F_{L/R}(s)$ away from the bound, so too can
$\alpha_{L/R}(t)$ be used to evolve the two accumulators towards their
respective bounds.  By setting both $\alpha_L(t)$ and $\alpha_R(t)$ to
positive values, each set of accumulators evolves towards their respective
bounds.  The alternative that is closer to the decision bound when this process
begins will reach its bound sooner, executing the corresponding response.

\subsection{Neural simulations}

\setlength{\figheight}{0.2\textheight}
\begin{figure*}
		\centering
		\begin{tabular}{lclc}
				\textbf{a} && \textbf{b}\\
				& \includegraphics[height=\figheight]{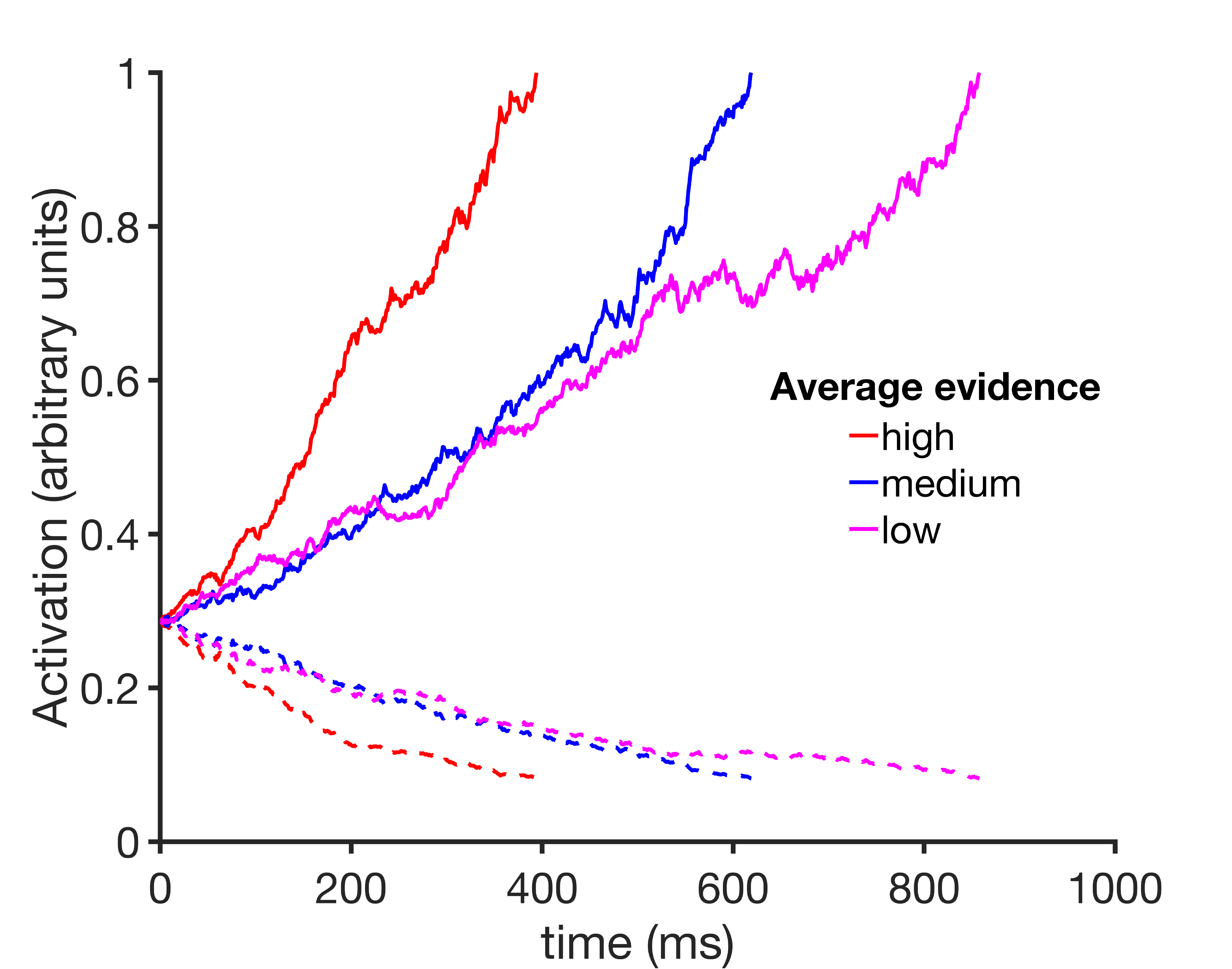}
					~
					\includegraphics[height=\figheight]{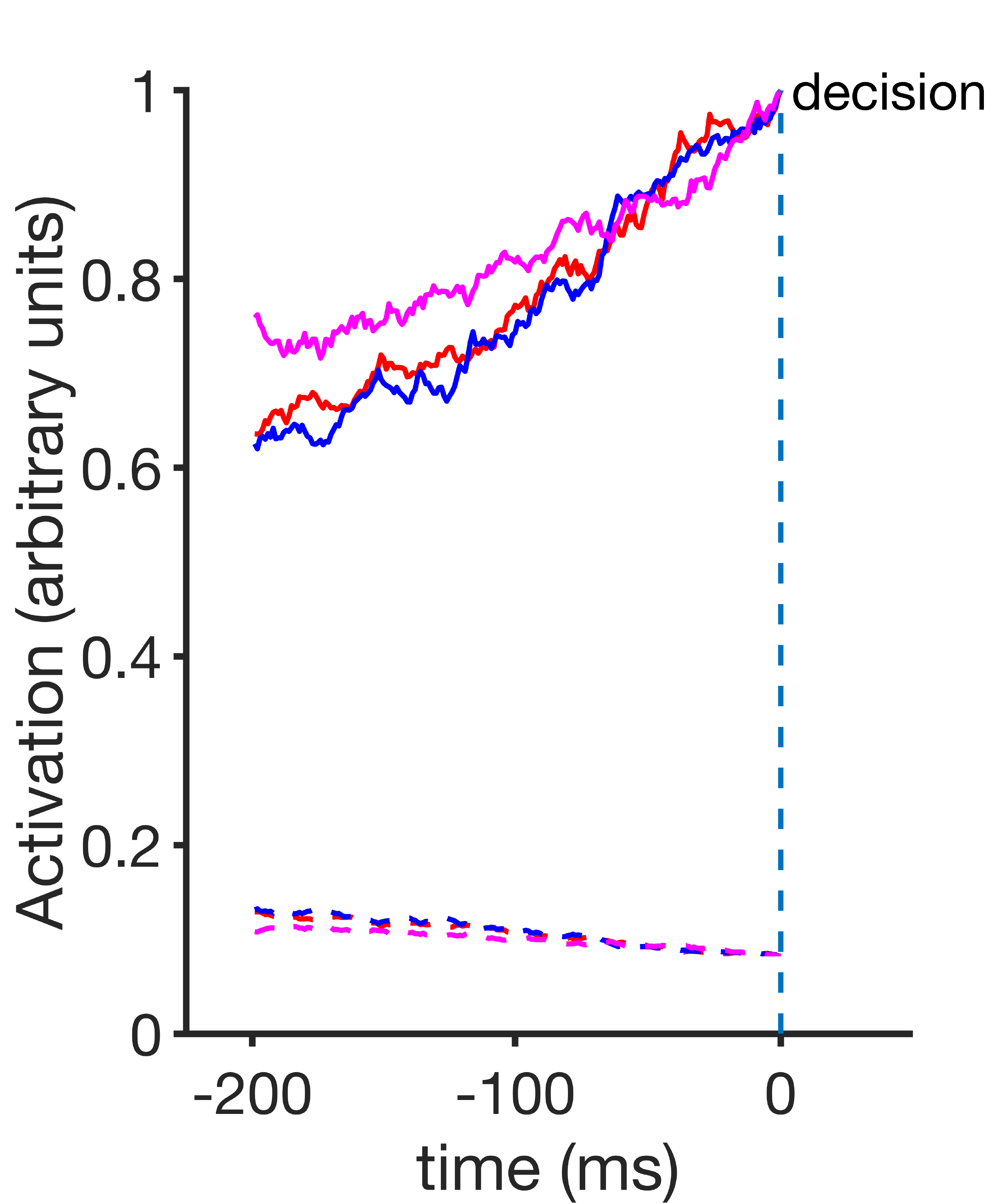}
				&&
				\includegraphics[height=\figheight]{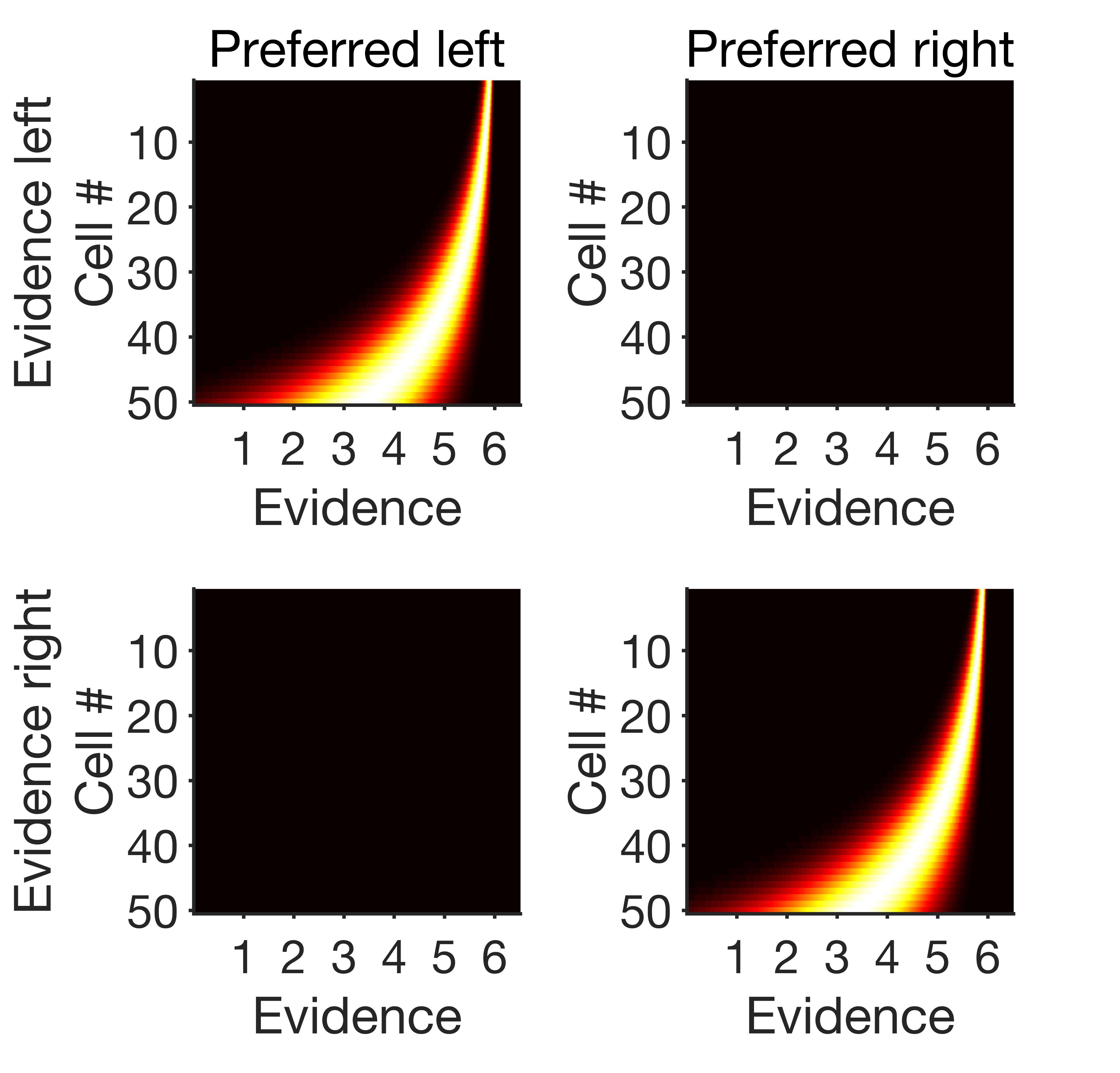}
		\end{tabular}
		\caption{\textbf{The Laplace transform and the inverse transform
		capture key findings from the neurobiology of evidence accumulation.}
		\textbf{a.}  The Laplace transform is shown for one value of $s$ in a
		simulated random dot motion experiment.  In this simulation,
$\alpha_L(t)$ was set at each moment to a normally-distributed random variable with a mean that
differed across condition.  In the solid curves, the mean was greater than
zero; in the dashed curves the mean was less than zero for $\alpha_L(t)$ (and
thus greater than zero for $\alpha_R(t)$). On the left we see that the firing
rate grew with time after information began to accumulate; the rate and
direction of accumulation depended on the mean of the random variable.  When
aligned to the time of response (right), firing rate showed a tight coupling.
If we changed the value of $s$, we would have observed the same qualitative
behavior, but with different rates of accumulation and more or less curvature
on average. Compare to Fig.~\ref{fig:Neural}a. 
\textbf{b.} Heatmaps for neurons participating in the inverse Laplace
transform.  Each line is the firing rate of a neuron. Neurons participating in
$\ftilde_L$ are shown on the top;  neurons participating in $\ftilde_R$ are
shown on the bottom.  On the left the firing rate is shown as a function of
evidence (confounded with time) when there is consistent evidence for the left
option.  On the right, the activation is shown when there is consistent
evidence for the right option.
Compare to Fig.~\ref{fig:Neural}b.
}
		\label{fig:modelneural}
\end{figure*}

In order to demonstrate that this approach leads to neural predictions that
are in line with known neurophysiological data, we provide simulations of two
paradigms that have received a great deal of attention in the
neurophysiological literature.  We first show that during the random dot
motion paradigm, units in $F(s)$ show activity that resemble classic
integrator neurons such as those observed in monkey LIP.  We then show that
during performance of a discrete evidence accumulation paradigm, units in
$\ftilde(\xstar)$ show receptive fields along the decision axis not unlike
those in mouse PPC.  
%
\subsubsection{Constant input in the random-dot motion paradigm}
We simulated the random dot motion paradigm by letting $\alpha(t)$ vary from
moment to moment by drawing from a normal distribution at each time step and setting
$\alpha_L(t) = -\alpha_R(t)$.  As in most applications of the diffusion model,
different levels of coherence were simulated by drawing from normal
distributions with different mean values,  for each of the levels
of coherence.\footnote{More precisely, we implemented $X_{t+\Delta} = X_t + A \Delta + c
\sqrt{\Delta} \mathcal{N}(0,1)$.   The value of $\Delta$ was set to $.001$,
$A$ was $1.25$, $.83$, and $.625$ across conditions. $X(0)=0$ and we
terminated the decision when $X=1$.}
Figure~\ref{fig:modelneural}a shows the result of
a representative unit with $s=2.5$ in $F(s)$.


Figure~\ref{fig:modelneural}a summarizes the results of these simulations.
Neurons in $F(s)$ grow towards a bound in a noisy way.  
When motion coherence is away from the neuron's preferred direction, the
firing rate 
instead decreases towards zero.
The rate of increase
is greater when there is more motion coherence (left), resulting in faster
responses for higher motion coherence.  When aligned to the time of response,
the unit's firing appears tightly aligned.  Although the response is actually
triggered by the activation of a particular unit in $\ftilde$, the activation
in $\ftilde$ is in one-to-one relationship with $F(s)$ (see
eq.~\ref{eq:LRftilde}) and all of the units in $F(s)$ corresponding to the
selected response monotonically increase as the decision bound becomes more
near.

Note that there is a slight curvature apparent in the model neuron shown in
Figure~\ref{fig:modelneural}a.  This curvature is a consequence of the
exponential function in the Laplace transform.  Choosing different values of
the $s$ would have resulted in more or less curvature.    Note the model
predicts that neurons encoding the inverse transform would show supported
receptive fields as a function of distance from the decision point.

\subsubsection{Information accumulators and position along a decision axis}

Figure~\ref{fig:modelneural}b shows analytic results from deterministic
evidence accumulation.  Analogous to the trials in the \citeA{MorcHarv16}
paper in which evidence was only presented in favor of one option or the
other, after initialization on left trials we set $\alpha_L(t)$ to a fixed
positive constant $X$ for the duration of the trial; $\alpha_R(t)$ was set to
$-X$.  On right trials the situation was reversed.  We show the activation of
units in $\ftilde_L(\xstar)$ and  $\ftilde_R(\xstar)$ on each type of trial.
The model captures the prominent features of the data. Units have receptive
fields along a subset of the decision axis. There are more units with
receptive fields near the time when the decision is reached.  The width of
receptive fields near the bound is less than the width of receptive fields
further from the bound.  Note that this is an analytic solution.  Had we
simulated detailed trials with sequences that included different amounts of
net evidence, the units would still have receptive fields along the net
evidence axis.

Note that neurons encoding the transform would grow or decay exponentially in
this task as a function of distance to the bound.   These neurons would have
the same relationship to the neurons in Figure~\ref{fig:modelneural}b as border
cells  \cite{SolsEtal08,CampEtal18} do to one-dimensional place
cells \cite{HowaEtal14}.  That is, in spatial navigation tasks, border cells
fire as if they encode an exponential function of distance to an environmental
border. If border cells manifest a spectrum of decay rates, then they encode
the Laplace transform of distance to that border. Taking the
inverse transform would result in cells with receptive fields that are
activated in a circumscribed region of space \cite{LeveEtal09}.

The results in Figure~\ref{fig:modelneural}b are closely analogous to
predictions for time cells and place cells from the same formalism
\cite{HowaEtal14,TigaEtal18a}.
Different sources of evidence trigger different sequences
leading to different decisions; this is closely analogous to experimental results
from time and place.  For instance, a recent study showed that distinct
stimuli in a working memory task trigger distinct sequences of time cells in
monkey lPFC \cite{TigaEtal18a}.  Similarly, distinct sequences of place cells
fire as an animal moves from a starting point along distinct trajectories
\cite{McNaEtal83}.


\subsection{Novel neural predictions}
The present approach is conceptually very different from previous
neural implementations of evidence accumulation models. In most prior models,  
the assumption is that many neurons provide  noisy exemplars of the decision
variable; an estimate of the decision variable can be extracted by averaging
over many neurons \cite<e.g.,>{ZandEtal14}.  
In contrast, although the activity of units in $F(s)$ are correlated with the
amount of instantaneous evidence, they are not noisy exemplars of that scalar
value.  Rather the Laplace transform  is written across many units with
different values of $s$.  At any moment, the activity of different units in
$F(s)$ are a known function of the decision variable, leading to quantitative
relationships both as a function of time for a particular unit but also
relationships between units.  Moreover, the model predicts specific
relationships between units in  $F(s)$ and $\ftilde(\xstar)$, which should
appear in pairs in decision-making tasks.
These properties  lead to several distinct neural predictions.  These
predictions are   summarized in Table~\ref{tab:neuralpredictions} and detailed
below.

\begin{table}
		\centering
		\begin{enumerate}
		\item Exponential functions for evidence accumulation.
				\label{pred:exp}
		\item Spectrum of rate constants $s$.
				\label{pred:spect}
		\item Laplace/Inverse pairs across tasks.				\label{pred:pairs}
		\item Linked non-uniform spectra of rate constants $s$ and $\xstar$.
				\label{pred:linked}
		\end{enumerate}
		\caption{Summary of novel neural predictions.  See text for details.}
		\label{tab:neuralpredictions}
\end{table}

\subsubsection{Prediction~\ref{pred:exp}: Exponential functions}
The hypothesis that the brain maintains an estimate of the Laplace transform
of the decision variable, rather than a linear function of the decision
variable leads to distinct predictions.
Even in circumstances where  evidence is accumulating at a constant rate, the
activity of integrator neurons should not in general be linear.  Note
that for values of $x$ small relative to $1/s$ it is difficult to distinguish
$e^{-sx}$ from $1-x$.  Moreover, the firing rate as a function of time should
be a function of the actual evidence as well as changing decision bounds.
However, systematic deviation from linearity is a clear prediction of the
present approach.

Note that a recent literature has questioned whether information accumulates
continuously in information accumulation neurons or whether it changes
abruptly in discrete steps \cite<e.g.,>{LatiEtal15}.  One could imagine that
the exponential function is expressed as a statistical average over a number
of units each of which obey step-like evolution following the same constant hazard
function.  It is also possible that work arguing for discrete steps in
activation has not considered the appropriate family of continuously-evolving
activation.  Indeed, recent work has begun to explore non-linear models for
activation as a function of time \cite{ZoltEtal18} and has found relatively
nuanced results.

\subsubsection{Prediction~\ref{pred:spect}: Spectrum of $s$ values}

The most fundamential prediction of the approach proposed here is that
different units coding for the decision variable integrate information with
different rate constants $s$.  We can think of each neuron's value of $1/s$ as
its ``evidence constant.''  The fundamental prediction of this approach is
that $s$ (and $1/s$) do not take on the same value across neurons.  
Although it has not to our knowledge been systematically studied, there is
evidence suggesting that neurons respond heterogeneously during simple
decision-making tasks.  For instance, Figure~7 in \citeA{PeixEtal18} appears
to show that different units discriminate the identity of a future response at different
points during a decision.  The computational approach proposed here requires
heterogeneity in $s$ values; a spectrum of $s$ values is essential to
construct an estimate of $x$ \emph{via} the Laplace domain. 

In most RDM experiments, the decision variable fluctuates in an unpredictable
way from moment-to-moment.  Moreover, changing decision bounds would also
change the firing rate for many neurons simultaneusly.   However, the
hypothesis that different units obey the same equations, but with different
values of $s$ suggests a strategy for estimating the spectrum.  Although
$\alpha(t)$ may fluctuate from moment to
moment based on the instantaneous evidence and perhaps also with changing
decision bounds, the same value of $\alpha(t)$ is distributed
to all of the neurons estimating $F(s)$ coding the distance to a particular bound. Thus
the momentary variability is shared across units.  Suppose we had two
simultaneously-recorded LIP neurons with the same receptive field and rate
constants $s_1$ and $s_2$.   On individual trials, each unit would appear to
fluctuate randomly over the trial as the net evidence fluctuates.  One unit
would move like a bead on a wire along the curve $e^{s_1 x(t)}$;  the other
would move like a bead on a wire along the curve $e^{s_2 x(t)}$.  Although the
fluctuations along each curve may appear random, plotting the firing rate of
one  neuron as a function of the other would result in an smooth exponential
curve.  In practice one would need to consider an appropriate time bin over
which to average spikes and undoubtedly face other technical issues in
performing this analyses.

\subsubsection{Prediction~\ref{pred:pairs}: Laplace/Inverse pairs across tasks}
In this paper we have noted that integrator neurons observed in RDM tasks have
properties analogous to the Laplace transform of distance to the bound,
$F(s)$, and that sequentially-activated cells in a virtual navigation decision
task have properties analogous to the inverse Laplace transform estimating
distance to the bound,
$\ftilde(\xstar)$.  The hypothesis presented here requires that the inverse
transform $\ftilde(\xstar)$  is constructed from the transform $F(s)$.
Moreover, the trigger for the actual response in the RDM task is the
activation of units with the smallest value of $\xstar$.  This approach thus
requires that both forms of representation should be present in the brain in
both tasks.

In tasks where instantaneous evidence is under experimental control and becomes
available at discrete times (such as the \cite{MorcHarv16} task), the Laplace
transform $F(s)$ should remain fixed during periods of time when no evidence
is available.  When evidence is provided, neurons coding for the transform
should gradually step up (or step down) their firing as the decision bound
they code for becomes closer (or further away).  Note also that one can
implement a leaky accumulator by including a non-zero value to $\alpha(t)$
during periods of time when no evidence is presented.   Conversely, in tasks
where the instantaneous evidence is not well controlled, such as the RDM task,
neurons coding for $\ftilde(\xstar)$ should still have ``receptive fields'' along
the decision axis, although in practice it may be much more difficult to
observe these receptive fields. 

Geometrically, $F(s)$ and $\ftilde(\xstar)$ have different properties that
might be possible to distinguish experimentally even in an RDM task where it
is difficult to measure single-cell receptive fields.  Consider a hypothetical
covariance matrix observed over many neurons representing either $F(s)$ or
$\ftilde(\xstar)$ during an RDM task in which the animal chooses
among two alternative responses \textsc{l} and \textsc{r}.    For units
representing $F(s)$, at the time of the response the population consists of
two populations.  That is, at the time of an \textsc{l} response,  neurons
participating in $F_L(s)$ are maximally activated and neurons participating in
$F_R(s)$ are deactivated.   This distinction will be reflected in the
covariance matrix; one would expect the first principal component of the
covariance matrix to correspond to this source of variance.  Now, for $F(s)$,
because neurons change their firing monotonically along the decision axis,
this principal component will also load on other parts of the decision axis,
changing smoothly.\footnote{Because the growth/decay of the units is not at
the same rate for each neuron, we would expect additional principal components
to capture this residual. }  In contrast, consider the set of neurons active
at the time of the response for a population representing $\ftilde(\xstar)$.
Like $F(s)$,  that the population of neurons in $\ftilde(\xstar)$ active at
the time of an \textsc{l} response is different than the population active at
the time of an \textsc{r} response.  However, unlike $F(s)$, because neurons
in $\ftilde(\xstar)$ have circumscribed receptive fields along the decision
axis, the neurons active at the time of the response for \textsc{l} are not
the same as the neurons active, say, three quarters of the way along the
decision axis towards an \textsc{l} response.  It may be possible to
distinguish these geometrical properties by examining the spectrum of the
covariance matrices.  

We will avoid commiting to predictions about what brain regions (or
populations) are responsible for coding $F(s)$ and which are responsible for
coding $\ftilde(\xstar)$ in any particular task.  It is clear that many brain
regions participate in even simple decisions \cite<e.g.,>{PeixEtal18}.
Moreover, the task in \citeA{MorcHarv16} is relatively complicated and 
may rely on very different cognitive mechanisms than the RDM task.    The
prediction is that there should be \emph{some} regions that encode the transform
and the inverse in any particular evidence accumulation task; it is possible
that these regions vary across tasks.

\subsubsection{Prediction~\ref{pred:linked}: Linked non-uniform spectra of rate constants $s$
and $\xstar$}
Experience with time cells and computational considerations
\cite{ShanHowa13,HowaShan18} leads to the  prediction that the spectra of
$s$ and $\xstar$ values should be non-uniform.  Although it may be challenging
to evaluate experimentally, Weber-Fechner scaling also leads to a
quantitatively precise prediction about the form of the distribution.

It is now clear that time cells are not uniformly distributed.  As a sequence
of time cells unfolds, there are
more neurons that fire early in the sequence and fewer that fire later in the
sequence.  This phenomenon can be seen clearly from the curvature of the
central ridge in Figure~\ref{fig:LaplaceNeural}b and has been 
evaluated statistically in a number of studies in many brain regions
\cite<e.g.,>{KrauEtal13,SalzEtal16,MellEtal15,JinEtal09,TigaEtal18a}.  If
similar rules govern the distribution of neurons coding for distance to a
decision bound we would expect to find more neurons with receptive fields near
the bound, with small values of $\xstar$, and fewer neurons with receptive
fields further from the bound with larger values of $\xstar$.  Weber-Fechner
scaling predicts further that the number of cells with a value of $\xstar <
\xstar_o$  should go up with $\log \xstar_o$, meaning that the probability
of finding a neuron with a value of $\xstar_o$ should go down like
$1/\xstar_o$.

The computational approach proposed in this paper links $F(s)$ and
$\ftilde(\xstar)$.  This naturally leads to the prediction that the
distribution of values for $\xstar$ and $s$ are linked.  Recalling that
$\xstar \equiv k/s$ enables us to extend the predictions about the
distribution of $\xstar$ values in populations representing $\ftilde(\xstar)$
to the distribution of $s$ values in populations representing $F(s)$.  There
should be a larger number of neurons with \emph{large} values of $s$ and fewer
neurons with \emph{small} values of $s$.  Weber-Fechner scaling predicts that the
number of neurons in $F(s)$ with values of $s > s_o$ should change linearly
with $\log s_o$.  Similarly,  the probability
of finding a neuron with a value of $s_o$ should go down like
$1/s_o$.

\section{Discussion}
The present paper describes a neural implementation of the diffusion model
\cite{Ratc78}.  Rather than individual neurons each providing a noisy estimate
of the decision variable, many neurons participate in a distributed
code representing distance to the decision bound.  Neurons coding for the
Laplace transform of this distance have properties that resemble those of
``integrator neurons'' in regions such as LIP.  Neurons estimating the
distance itself, constructed from approximating the inverse transform, have
receptive fields along the decision axis.

\subsection{Alternative algorithmic evidence accumulation models in the
Laplace domain}

In this paper we have focused on implementation of the diffusion model.
However, it is straightforward to formulate other algorithmic models in the
Laplace domain.  For instance, in order to implement a race model, all that
needs to be done is to make $\alpha_L$ and $\alpha_R$ independent of one
another rather than anticorrelated as in the present implementation of the
diffusion model.  Similarly, one can implement a leak term
\cite{BuseTown93,UsheMcCl01,BogaEtal06} by including a constant negative term to 
each of the $\alpha(t)$ values.   The magnitude of the constant term is
proportional to the leak parameter.

Insofar as these approaches are equivalent to the corresponding algorithmic
model, adopting this form for neural implementation entails at most subtle
behavioral distinctions to classic implementations of the diffusion model.
However, there is one unique property that follows
from this approach of formulating decision-making as movement along a
Weber-Fechner compressed decision axis.  As discussed above, the logarithmic
compression along the $\xstar$ axis means that the neural representation is
not equidiscriminable in the decision variable $X$, but in $e^X$.  If $X$ is
understandable as the log likelihood ratio (Eq.~\ref{eq:seqdeclog}), then this
means that the magnitude of the change in the likelihood ratio itself 
controls the change in the neural representation.  This would predict that the
just-noticeable-differences in the decision variable are controlled by the
likelihood ratio itself rather than the log-likelihood ratio.  In practice
this prediciton would be difficult to assess in any particular experiment
because in sequential decision-making tasks the likelihood ratio as a function
of time is not a physical observable but is inferred based on assumptions
about the processing capabilities of the observer.

%

\subsection{Unification of working memory, timing and evidence accumulation in
the Laplace domain}

As mentioned in the introduction, many of the most influential ideas in
computational neuroscience are derived from cognitive models developed without
consideration of neural data.   Often in cognitive psychology these models are
treated as independent of one another.  One of the contributions of the
present paper is that it enables the formulation of cognitive models of evidence
accumulation in the same Laplace-transform formalism as cognitive
models of other tasks, including a range of memory and timing studies
\cite{ShanHowa12,HowaEtal15,SingEtal18}.  This paper also enables
consideration of neural models of evidence accumulation in the same formalism
as neural models of representations of time and space
\cite{HowaEtal14,TigaEtal18a}.

Although in  cognitive psychology, models of working memory and evidence
accumulation have been treated as largely separate topics \cite<but
see>{UsheMcCl01,DaveEtal05}, in computational neuroscience they have been
treated in much closer proximity.  Indeed, the state of a perfect integrator
reflects its previous inputs (Eq.~\ref{eq:DiffusionDef}).  In computational
neuroscience, decision-making models use similar recurrent dynamics
\cite{WongEtal07,Wang02} that are  used for working memory maintenance
\cite<e.g.,>{ChauFiet16,MachEtal05,RomoEtal99}.  At the neural level, a
perfect integrator predicts stable states in the absence of external
activation \cite{FunaEtal89}, analogous to the binary presence of an item in a
fixed-capacity working memory buffer.

In cognitive psychology models of evidence accumulation and timing have been
more closely linked.  Noting that in the case of a constant drift rate, $X_t$
is proportional to the time since evidence accumulation began, models have
used integrators to model timing
\cite{RiveBeng11,SimeEtal11,SimeEtal13,LuzaEtal16,BalcSime16}.  ``Diffusion models for
timing''  are also closely related to models of
learning \cite{GallGibb00,KillFett88,LuzaEtal17}  and the scalar expectancy
theory of timing \cite{GibbChur84,GibbEtal84}.  At the neural level, these
models predict that during timing tasks, firing rate should grow monotonically, a phenomenon that has been reported 
 \cite<e.g.,>{KimEtal13,XuEtal14}.

In both of these cases, the neural
implementations of the cognitive models has found some neurophysiological support.   
In each of the foregoing cases, neural models implement a scalar value from an
algorithmic cognitive model as an average over many representative neurons. In
the case of timing models as in evidence accumulation models, this predicts
monotonically increasing or decreasing firing rates.
As we have seen here, monotonically increasing or decreasing functions can be
generated by neurons participating in the Laplace transform of a function
representing a scalar value over a set of neurons.  However, the present
framework also predicts that populations with monotonically changing firing rates
should be coupled to populations with sequentially activated neurons like
those found in $\ftilde(\xstar)$ and that these two forms of representation
are closely connected to one another.  Thus, findings that working memory
maintenance is accompanied by sequentially-activated cells during the
delay period of a working memory task \cite{TigaEtal18a} or that timing tasks
give rise to sequentially activated neurons \cite{TigaEtal17b} can be readily
reconciled with the present approach but are difficult to reconcile with
models that require representation of a scalar value as an average over many
representative neurons.  The present approach and much previous work in
computational neuroscience differ in their account of  how
information is distributed over neurons.   

Function representation \emph{via} the Laplace domain also has potentially
great explanatory power in developing neural models of relatively complex
computations.  It has been long appreciated that computations can be
efficiently performed in the Laplace domain.  In much the same way that
understanding of the mathematics lets us readily specify how to set $\alpha$
to accomplish various goals such as initializing with a specific response bias
in the evidence accumulation framework described here, access to the Laplace
domain lets us readily write down mechanisms for translation of functions
\cite{ShanEtal16} or perform arithmetic operations on different functions
\cite{HowaEtal15}.  Computations in the Laplace domain require a
large-scale organization for macroscopic numbers of neurons;  this
large-scale organization enables basic computations.  Insofar as many
different forms of information in the brain use the same form of compact
compressed coding, one can in principle recycle the same computational
mechanisms for very different forms of information, including sensory
representations, representations of time and space, or numerosity.

\bibliography{/Users/marc/doc/bibdesk}

\end{document}

Can these computational models from mathematical psychology be unified? 
If one could do so, and understand the mapping between the algorithmic level
and neural implementational level supporting the formal model, then
one would have a formal framework in which to construct cognitive models
at the algorithmic level and map them onto neural circuits.
As a step towards this goal, we extend a formalism based on the Laplace
transform.  This formalism has been developed
to account for cognitive models in memory and applied to neural 
In this paper we extend a formalism that has been applied to representations
of time and space in support of models of memory of various forms and extend
it to